%% file: main.tex
\documentclass{article}

\usepackage{microtype}
\usepackage{graphicx}
\usepackage{subfigure}
\usepackage{booktabs} 

\usepackage[hyphens]{url}
\usepackage{hyperref}
\usepackage{multicol} 
\usepackage{url}
\usepackage{booktabs}
\usepackage{tabularx}
\usepackage{enumitem} 
\usepackage{multirow} 
\usepackage{caption}   
\usepackage{float}
\usepackage{graphicx} 
\usepackage{framed}   

\newlist{inlinelist}{enumerate*}{1}
\setlist[inlinelist,1]{label=(\arabic*), before=\unskip{: }, itemjoin={{; }}}

\newcommand{\mfont}[1]{{\textsf{\textsc{#1}}}}

\newcommand{\codegemma}{\mfont{CodeGemma-7b}}
\newcommand{\codellama}{\mfont{CodeLlama-13b}}
\newcommand{\llamasm}{\mfont{Llama 3.1-8b}}
\newcommand{\llamamd}{\mfont{Llama 3.1-70b}}
\newcommand{\zeroone}{\mfont{Yi-Coder-9b}}
\newcommand{\codestral}{\mfont{Codestral-22b}}
\newcommand{\starcoder}{\mfont{StarCoder2-15b}}
\newcommand{\mistral}{\mfont{Mistral-7b}}
\newcommand{\gpto}{\mfont{GPT-4o}}
\newcommand{\gptm}{\mfont{GPT-4o-mini}}
\newcommand{\magicoder}{\mfont{Magicoder 6.7B}}
\newcommand{\gemini}{\mfont{Gemini 1.5-F}}
\newcommand{\claude}{\mfont{Claude 3.5-S}}
\newcommand{\qwen}{\mfont{Qwen 2.5CI-14b}}

\usepackage{listings}

\usepackage{mdframed}

\usepackage[table,dvipsnames]{xcolor}

\definecolor{pos}{HTML}{2E8B57} 
\definecolor{neupos}{HTML}{FFA500}    
\definecolor{neuneg}{HTML}{FA8072}    
\definecolor{neg}{HTML}{B22222} 
\usepackage{tcolorbox}

\usepackage[accepted]{icml2025}

\usepackage{amsmath}
\usepackage{amssymb}
\usepackage{mathtools}
\usepackage{amsthm}
\usepackage{bbm}

\usepackage[capitalize,noabbrev]{cleveref}

\theoremstyle{plain}

\theoremstyle{definition}

\theoremstyle{remark}

\usepackage[textsize=tiny]{todonotes}

\icmltitlerunning{CLOVER: A Test Case Generation Benchmark with Coverage, Long-Context, and Verification}

\begin{document}

\twocolumn[
\icmltitle{CLOVER: A Test Case Generation Benchmark \\ with Coverage, Long-Context, and Verification}



\icmlsetsymbol{equal}{*}

\begin{icmlauthorlist}
\icmlauthor{Jiacheng Xu}{}
\icmlauthor{Bo Pang}{}
\icmlauthor{Jin Qu}{}
\icmlauthor{Hiroaki Hayashi}{}
\icmlauthor{Caiming Xiong}{}
\icmlauthor{Yingbo Zhou}{}
\\
\textnormal{ Salesforce AI Research }

\end{icmlauthorlist}


\icmlcorrespondingauthor{Jiacheng Xu}{jiacheng.xu@salesforce.com}

\icmlkeywords{Machine Learning, ICML}

\vskip 0.3in
]




\begin{abstract}

Software testing is a critical aspect of software development, yet generating test cases remains a routine task for engineers. This paper presents a benchmark, CLOVER, to evaluate models' capabilities in generating and completing test cases under specific conditions. Spanning from simple assertion completions to writing test cases that cover specific code blocks across multiple files, these tasks are based on 12 python repositories, analyzing 845 problems with context lengths ranging from 4k to 128k tokens. Utilizing code testing frameworks, we propose a method to construct retrieval contexts using coverage information. While models exhibit comparable performance with short contexts, notable differences emerge with 16k contexts. Notably, models like GPT-4o and Claude 3.5 can effectively leverage relevant snippets; however, all models score below 35\% on the complex Task III, even with the oracle context provided, underscoring the benchmark's significance and the potential for model improvement.
The benchmark is containerized for code execution across tasks, and we will release the code, data, and construction methodologies.

\end{abstract}

\input{tab_related_work_comp}

\section{Introduction}

Software testing is integral to the software development lifecycle \cite{yoo2012regression, wang2024software, alshahwan2024automated}. From test-driven development \citep{mathews2024testdrivendevelopmentcodegeneration} to program repair \cite{yasunaga21a,swebench}, crafting efficient and high-quality test cases is a routine task. Recently, large language models (LLMs) have gained attention for their potential in code and software testing enhancements. These models utilize context, user prompts, history, and code prefixes for code suggestions \cite{CodeGen,yicoder,codellama,qwen2.5}. 
To evaluate models' capability in writing code, many benchmarks have been proposed in the past few years. 
These benchmarks vary in focus, tackling areas such as basic coding problems \cite{mbpp, chen2021evaluating}, data science tasks \cite{Lai2022DS1000}, reasoning challenges \cite{crux}, and issue resolution \cite{swebench}. We summarize the most relevant work in Table~\ref{tb:related}. 

\textit{Can LLMs write executable test cases with specific requirements in a realistic setup?}
To address this question, we create a benchmark, CLoVer, focusing on automatic evaluation of unit test case generated by LLMs.  
We create an automatic pipeline with little human intervention to scrape permissive repositories from GitHub and configure the execution environment. 
We identify potential problems by extracting and verifying test cases from the existing codebase. After this step, we take these problems and  structure three challenging tasks: (1) \textit{Task I} simulates a code completion tool by focusing on cloze-filling style questions; (2) \textit{Task II} addresses scenarios requiring coverage and testing of specific methods or classes within the source code; (3) \textit{Task III} involves improving code coverage, where models are challenged to cover certain code blocks within the source.
The selection of example is driven by AST parser and code coverage results.
We evaluate model performance by executing the generated code and capturing line coverage, offering a tangible measure of their effectiveness.



In practical software testing, leveraging a comprehensive context window is crucial, encompassing dependencies and their antecedents. To evaluate models in a realistic context-aware fashion, we construct \textit{oracle context} via test coverage for each example. We assess model performance across three tasks with context lengths spanning 4k to 128k tokens and introduce \textit{context utilization} as a metric to assess how effectively models leverage extended contexts, independent of their absolute performance.

Our evaluation includes 10 open-source and 4 proprietary models. In Task I, many open-source models, such as \mistral{} and \qwen{}, underperform with longer contexts, indicating a decline in response quality despite their technical capacity to handle such lengths. 
In Tasks II and III, all models encounter difficulties in generating executable code, even when provided with oracle context. A notable trend is the sharp performance drop among open-source models starting at a 16k context window. The highest performance across all tasks is demonstrated by \claude{} and \gpto{}, with \gpto{} achieving a 32.7\% success rate on the most demanding task, Task III. 
\textit{We identified a significant disparity in context utilization and long-context instruction-following capabilities between leading proprietary models and others.
}
Our data pipeline and evaluation sandbox are designed for scalability. We plan to release the code, benchmark, Dockerized environment, and recipes to enable the community to use these resources for further development and training. The benchmark also supports code agent by providing APIs and task instructions.

\begin{figure*}[t]
\begin{center}
\small
\centerline{\includegraphics[width=\textwidth]{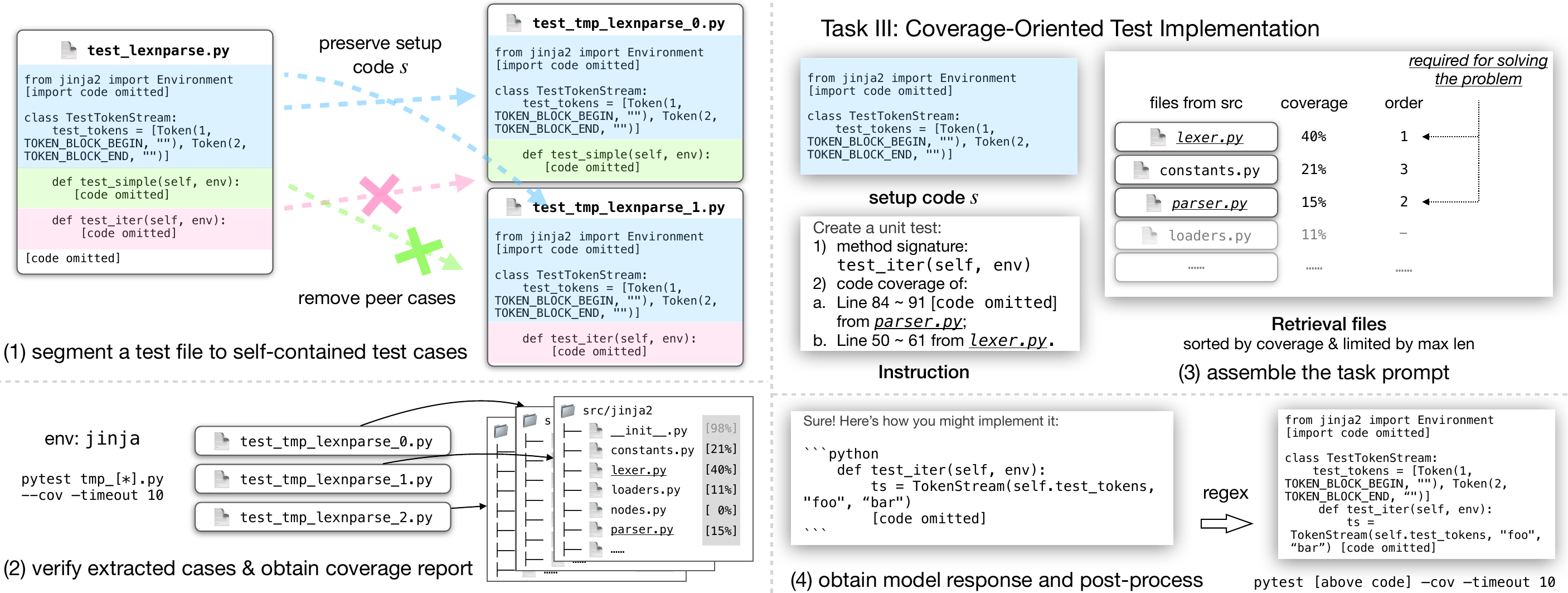}}
\caption{Pipeline overview. In this example, we focus on a test function \texttt{test\_iter}, which covers the use of \texttt{Token} and \texttt{TokenStream} classes from the source code. There are four major steps \begin{inlinelist}
    \item  we extract the problem from a test file \texttt{test\_lexnparse.py}.
    \item  verify of the extracted case(s) by running pytest
    \item assemble task prompts with pre-constructed oracle dependent files
    \item obtain model response and verify the execution status 
\end{inlinelist}  In Task I, we mask part of the assertion statements. In Task II and III, we ask model to complete the test code almost from scratch with constraints imposed. }
\label{fig1}
\end{center}
\end{figure*}

\input{tab_task_stat}

\section{Data \& Sandbox Construction}
\label{sec:setup}

In Figure~\ref{fig1}, we describe the overall pipeline from data collection to final evaluation. In this section, we will primarily focus on data collection and environment setup. 

\subsection{Data Collection}

\textbf{Source identification}\quad 
Following \cite{swebench}, we began by scraping 42 new python repositories and ultimately narrowed it down to 12 repositories for actual use. Details and reasons for exclusions are provided in Appendix \ref{app:sec:repo}. In our methodology, we identified folders containing source and test code by matching filenames with the pattern \texttt{test\_*.py}. For eleven repositories, we could not extract test suites. This process resulted in the identification of test modules, each comprising at least one test function.

\textbf{Problem extraction from file} \quad
Test cases are extracted from modules by parsing Python files using the \textsc{ast} tool to identify test functions. Using heuristics, we isolate setup code $s$ to remove unrelated test functions. In Figure~\ref{fig1}, test functions \texttt{test\_simple} and \texttt{test\_iter} are preserved with the necessary setup code, resulting in self-contained problems named \texttt{tmp\_test\_lexnparse\_[*].py}. We maintain the original structure and path of test modules.

\textbf{Verification API \texttt{verify}} \quad
Executing new unit tests requires careful design. During the design and testing of our \texttt{verify} API, we considered several points: (1) Consistency check. Evaluate model-generated implementations against ground-truth code to identify issues from extraction, heuristics, or system conditions such as caching;
(2) Batchify operations. Enable batch evaluation of test cases to decrease overhead from test framework executions and setups; 
(3) Timeout management. Prevent infinite loops in model-generated code;
(4) Error handling and logging;
(5) Repository restoration. Ensure repository state is reset before and after each use.
We wrap this verification process to an API  $\texttt{verify}(case) \rightarrow  \{\mathrm{true}, \mathrm{false}\} $ where the output indicates whether the $case$ can execute successfully.

\textbf{Coverage API \texttt{cov}}\quad
The coverage API provides line coverage metrics for a test case across the entire repository. Utilizing pytest-cov, it reports hit and missed lines, and computes a file-level coverage rate, even if execution fails. Unlike \texttt{verify}, \texttt{cov} cannot be parallelized due to shared cache dependencies but can still deliver coverage reports on failed tests.

\subsection{Sandbox Construction}
To run test programs across different repositories, we create sandboxes and package them in a Docker image, maintaining minimal intervention to ensure the process is scalable to a larger number of repositories.

\textbf{Procedure}\quad
First, we create a conda virtual environment with Python version set to 3.10. Then we install packages including \texttt{poetry}\footnote{\url{https://python-poetry.org/}} and \texttt{tox}\footnote{\url{https://tox.wiki/en/stable/}}. We exhaustively search for \texttt{txt} files, and try to \texttt{pip install} those files. Then, \texttt{git submodule} related operations will handle submodules under the project if any. After this step, we try to install the package from the current project directory with \texttt{pip}, \texttt{poetry} and \texttt{tox}. After all the steps, we run \texttt{pytest} to check if we can find a significant number of passed test cases. 
In practice, the procedure above can automatically configure the environment of 25 out of 42 (59.5\%) repositories.
We describe more detail about construction failure in Section~\ref{app:sec:repo}.

\textbf{Efficiency}\quad Tasks are evaluated sequentially, while evaluations within each task run concurrently across different repositories. The longest-running repository determines the evaluation's time bottleneck. To limit evaluation to 2 hours per model on a CPU Linux machine, we capped the maximum number of examples per repository: 50 for Task I, and 25 each for Task II and III.

\subsection{Evaluated Models}
We utilized vLLM \cite{vllm} for model inference with temperature and top\_p set to $0.2$ and $1.0$. 
Maximum output lengths were 200, 4,000, and 4,000 for Tasks I, II, and III, respectively. To accommodate output tokens without exceeding model length limits, we adjusted the maximum sequence length by the output length during data preparation. The tokenizer from \mistral{} was used in this process. 
We evaluated open-source models including \codegemma{} \citep{team2024codegemma}, \magicoder{} \cite{wei2024magicoder}, \qwen{} (Coder-Instruct) \cite{qwen2.5},  \zeroone{} \cite{yicoder}, \starcoder{} \cite{starcoder2}, \codellama{} \citep{codellama}, \llamasm, \llamamd{} \citep{dubey2024llama}, \codestral, and \mistral{} \citep{jiang2023mistral}.
For proprietary models, we evaluated \claude(onnet), \gemini(lash), \gpto{} (2024-08-06), and \gptm{} (2024-07-18). 

\section{Construction of Oracle Retrieval}
To write or complete test cases, models need access to the related source code. To offer a simplified but \textit{realistic} evaluation setting without using agents or retrievers, we provide oracle retrieval code in this benchmark. 
This leverages our executable environment and the \texttt{coverage} API for detailed coverage information. This setup aims to: (1) explore models' near-upper bound performance, (2) and test models in long-context scenarios.
Our approach constructs long-contexts naturally and demands a multi-hop understanding of code and effective information use.
Our setup is also perfect for software agent development.

\textbf{Motivation}\quad Files such as \texttt{\_\_init\_\_.py} are often highly covered by most test cases, but they contribute little value in terms of addressing specific problems, as per information theory. These files can quickly deplete the context budget due to their high coverage rates. Hence, we need to calibrate the coverage information to reflect the importance of certain informative files. 
\input{fig2}
\textbf{Objective}\quad The objective of constructing the oracle retrieval is to provide the most relevant or informative content within a constrained context budget. For the rest of this section, we will describe how we prioritize salient information with coverage information. 

\subsection{Calibration of Coverage}
Within a file, we represent the test cases as \( Y = \{ y_1, y_2, \ldots, y_T\} \). Typically, these cases share some setup code and are organized under the same testing topic. 
The collection of all source files is denoted as \( X = \{x_1, x_2, \ldots, x_F\} \). 
When using the \texttt{verify} API on \( T \), we get a coverage tensor \(\mathbb{C} \in \{1,0\}^{T \times F \times L} \), where \( C_{t,f,l} = 1 \) indicates test case \( y_t \) covers the \( l\)-th line of file \( x_f \), and \( C_{t,f,l} = 0 \) otherwise. $T  = |Y|$ represents the total number of test cases in this file, $F = |X|$ is the total number of source files, and $L$ is the max number of lines in $src$. 
We run pytest-cov two times for each test case $y_t$:
\begin{itemize}[noitemsep, topsep=1pt, partopsep=0pt,left=2pt]
    \item a regular run $C^{base}_t$. This will return the regular coverage report of $y_t$ over $X$.
    \item empty run $C^{empty}_t$. In this setting, we replace the code with an empty test statement: \texttt{def test(): assert True} and it will be deployed to the same location of $y_t$. For instance, if \texttt{test\_iter} was implemented in \texttt{tests/util}, we will deploy the empty test to that directory as well. 
\end{itemize}

\textbf{Repository Baseline}\quad
We propose a repository baseline is established by comparing $C^{base}_t$ and $C^{empty}_t$. 

\begin{equation*}
    Q_t^{\text{repo}} = \left\{ x_f \mid    arg_f [ \mathbbm{1}(C_{t,f}^{base} = C^{\text{empty}}_{t,f})  ] \right\}
\end{equation*}

where $\mathbbm{1}(\cdot)$ is the indicator function.
The set \( Q_t^{\text{repo}} \) comprises the files \( x_f \) for which, for any test case index \( t \), the coverage remains unchanged after executing the actual test case. This implies that the files in \( Q_t^{\text{repo}} \) offer minimal information gain in terms of entropy for generating test case \( y_t \).

\textbf{Peer Baseline}\quad
To uniquely identify each test case, we set a Peer Baseline. The aim is to identify the most distinctive information across test cases. For a particular test case \(y_t \), the Peer Baseline is defined as follows

\begin{equation*}
    Q_{t}^{\text{peer}} = \left\{ 
      x_f \mid arg_f [
        \mathbbm{1}(\sum_{t'=1}^T C_{t',f,l}^{base} = 1)
      ]
    \right\}
\end{equation*}

where $\mathbbm{1}(\cdot)$ is the indicator function.
\(\sum_{t'=1}^{T} C_{t',f,l}^{base} = 1\) ensures that the line \( l \) is covered by exactly one test case (test case \( y_t \)), meaning it's only covered by the test case \( y_t \). 
Next, we define \( Q' \), which is the set not meeting the criteria for either \( Q^{\text{peer}} \) or \( Q^{\text{repo}} \): $Q'_{t} = F \setminus \left( Q_t^{\text{repo}} \cup Q_t^{\text{peer}} \right)$.
We regard the value of files in \( Q' \) as lower than those in \( Q^{\text{peer}} \) but higher than the repository baseline \( Q^{\text{repo}} \).

\textbf{Calibration of Coverage}\quad
Source files are classified into three categories: \( Q^{\text{repo}} \), \( Q^{\text{peer}} \), and \( Q' \).
The approach for assembling context for test case \( t \) gives precedence to files in the order of \( Q_t^{\text{peer}} \), followed by \( Q_t' \), and ultimately \( Q_t^{\text{repo}} \).
Within each category, we randomly select files if the context budget does not permit using them all.


\subsection{Task Setup}
Before diving into these three tasks, we define some terminologies which share across these tasks.
For one task instance, we provide three categories of contents:
\begin{itemize}[noitemsep, topsep=1pt, partopsep=0pt,left=2pt]

    \item Task instruction. We show  examples in Fig~\ref{fig1} and \ref{fig:prompt_box}.
    \item In-file code, including setup code $s$ and function declaration $f$. Setup $s$ prepares necessary components, such as imports, fixtures, and any initial configurations, required for the test. Function declaration $f$ specifies the function's name, arguments, and any return types, if applicable. In Task I, we also provide code prefix, which will be discussed later.
    \item Source files per task requirement and from oracle retrieval. Files required by task are guaranteed to be provided unless in the Problem Only setting. It also has higher priority compared to oracle retrieval when we try to fill the context budget.
\end{itemize}

\textbf{Setting} \quad We introduced two settings across three tasks, Problem Only (PO) and Contextual. In Problem Only setting, we only provide the Task instruction and in-file code. In contextual setting, we provide code snippets capped by context budget. 

\section{Task I: Mask Prediction in Assertion Statements}
This task challenges the model to predict the missing element in an assertion statement within a test case, addressing the code completion feature offered by coding companions.


\textbf{Problem Formulation}\quad
For each problem $x$, it has following elements in the prompt of the task $[s, f, p, q, ref]$ in the Problem-Only setting:

\input{tab_task1_agg}

\begin{itemize}[noitemsep, topsep=1pt, partopsep=0pt,left=2pt]

    \item Prefix ($p$) refers to the existing code within the test function, serving as the context for solving assertion statements.
    \item Assertion statement with a \texttt{MASK} ($q$) represents the task for models to complete. Based on the surrounding code and any referenced materials, the model is expected to fill the gap and complete the assertion statements. $q$ contains exactly one \texttt{MASK}. 
    \item Reference answer ($ref$) is the original answer from the code. Any valid answer passing RER, a metric defined next, is acceptable as correct.
\end{itemize}
The model relies solely on the problem details, without extensive information about the method under test.

\textbf{Cloze construction}\quad
AST tools identify all possible assertion statements, including unary (e.g., variables, functions) and binary operations (comparisons). For binary, either operand can be masked. The suffix of \( q \) is removed to avoid hints, ensuring \( q \) is the last code line.

\textbf{Example selection}\quad
Preliminary study find that within each repository, there exists certain high frequent assertion statements, which provides unwanted hint to models. For instance, ``200'' (string literal) and 200 (number) are the most frequent candidates for the \texttt{MASK}. 
So we filter out problems with common $ref$. The chosen probability of of a problem $x_i$ is defined as: 
\begin{equation*}
    p(x_i) = 
    \begin{cases} 
      0, & \text{if} \quad \frac{\text{count}(ref_i)}{N} > 0.01 \\ 
      \frac{\text{len}(ref_i)}{ \sum_{j=0}^{N} \text{len}(ref_j)}, & \text{otherwise}
    \end{cases}
\end{equation*}
where $N$ is the total number of problems in one repository.
We downsample to 50 problems per repository to maintain a diverse set of problems.

\textbf{Prompt Template} \quad
We explore two elicitation methods \begin{inlinelist}
\item answer only ($pred$): the model yields the answer directly in a code block
\item assertion statement with answer filled \( q.\texttt{replace}(\texttt{MASK}, pred) \): the model returns the line with blank filled. 
\end{inlinelist}
Our studies with \codellama{}, \mistral{}, \codegemma{}, and \codestral{} show the filled assertion method improves execution rate by at least 6.0\%, thus we use it for Task I experiments.

\textbf{Metrics \& Verification} \quad
Let the model prediction for \texttt{MASK} be $pred$.
We implement three evaluation metrics for this task
\begin{inlinelist}
    \item Exact match is defined as $\mathrm{EM} = \mathbf{1}(ref = pred)$
    \item Execution Rate (ER) indicates the execution result of the assertion statements filled with $\text{pred}$
    \item Refined execution rate (RER) is based on ER but we applied post-processing steps to remove trivial assertions and prohibit the copy of an existing assertion from the context. 
\end{inlinelist}

Post-processing discards the following invalid scenarios \begin{inlinelist}
\item $pred$ is a constant in a unary operation
\item $pred$ is a constant in a binary operation where the other operand is also a constant
\item in an equation comparison, \(pred\) matches the operand on the opposite side.
\end{inlinelist}
We follow the definition of constant in AST. 
Since the problems are selected given its surface length, as a proxy to its difficulty, the false negative ratio is considered low.

\input{tab_task1_ctx}

\subsection{Results in the Problem Only setting} In this setting, we only provide the problem itself, excluding external context like the code of MUT. 
We present the result in Table~\ref{tb:task1}. The best open-source model in this setting is \qwen{}, achieving comparable performance compared to proprietary models. \claude{} performs the best in all metrics. 

\input{tab_task2}
\textbf{Gap between EM and (R)ER} \quad
We analyzed instances where predictions were successful in terms of RER but failed under the exact match (EM) criteria. The model's predictions averaged 25.8 characters compared to 30.7 characters in ground truth answers, suggesting a tendency toward brevity or shortcuts. Common scenarios where execution succeeds but predictions do not exactly match the original code include
\begin{inlinelist}
    \item Unary operations in \( q \) with non-constant \( \text{pred} \) (e.g., \texttt{assert} \( \text{pred} \))
    \item Use of syntactic sugar, such as considering \texttt{x in dict} equivalent to \texttt{x in dict.keys()}
    \item String manipulations, including \texttt{strip} functions and interchangeable quotation marks
    \item Assertions for non-existence, like \texttt{assert x not in y}, where \( x \) is flexible
\end{inlinelist}
Currently, the execution method lacks understanding of contextual semantics or user intent. A future goal is to develop a tool that can evaluate responses based on execution success and alignment with user intent.

\subsection{Results in Contextual Settings}
We present the result in Table~\ref{tb:task1ctx}. 
The best overall performance is achieved by \claude{} in both settings, with a 3.0\% gain from 72.6\% to 75.6\%.
In the contextual setting, we found there is a sharp decrease after 8k max length in most open source models including \codellama{}, \starcoder{}, \mistral{}, \qwen{}, and \codestral{}. We examined the model response in these cases and we find that the chance of getting gibberish output increases along with the increase of input length. Note that the prompt template remains the same for context free and contextual setting where the only change applied is the additional code snippets.

\textbf{Context Utilization $\Delta$}\quad
We introduce a novel metric to measure models' capability in effectively utilizing the context. On the performance gain side, we define it as
$\Delta_{max} = \max(r_{\text{4k}}, r_{\text{8k}}, \ldots, r_{\text{maxLen}}) - r_0$ where $r_0$ is the context free baseline performance. Since we provide oracle context to the model, shorter context carrying strong hint could be sufficient and even better than longer sequence. $\Delta_{max}$ measures the best possible gain a model could get. Vice versa, we define $\Delta_{min} = \min(r_{\text{4k}}, r_{\text{8k}}, \ldots, r_{\text{maxLen}}) - r_0$. This set of metrics focuses on the relative performance change given longer context. The ideal value, if context provides good source of information, for this metric follows this equation $\Delta_{max} > \Delta_{min} > 0$.

\section{Task II: Targeted Test Implementation}
In Task II and Task III, we will shift from code completion to open-ended code generation, which is more challenging and requires longer context. 
In Task II, given a python class or function from the source code, the model needs to complete the test code by using the target.

\textbf{Problem Formulation}\quad
For each problem, we provide setup $s$, function declaration $f$ and a specification to use the target. We show the specification template and an example in Figure~\ref{fig:prompt_box}.
The ``target\_name'' here is the name of the class or function. The ``type'' is either ``class'' or ``function''. ``file\_name'' is where the target object was implemented. 

\textbf{Data Construction}\quad
We use AST tools to parse the code and identify all \texttt{Attribute} type nodes through a recursive walk to find suitable targets. These identified classes and functions become potential targets. They are then matched with those covered by this case in \(C_t^{base} \). A random target is selected as the requirement. In settings ranging from 8k to 64k, we ensure the inclusion of the necessary file as specified. The maximum length constraint is 8k and the output length is 4k, thus the combined length of instructions and the required file must not exceed 4k. Any cases exceeding this limit are discarded. By setting a single target, we maximize the inclusion of examples.

\textbf{Answer Format}\quad
In this task, the generated code is the completion of the function declaration $f$. 
We designed two prompt templates to capture the output, one with the completion part only ($prompt_{part}$), and one with the full code block ($s, f$) along with the completion $prompt_{full}$. 

\textbf{Metrics}\quad
We define two metrics for Task II and Task III. 
\textit{Execution Rate} measures if the generated code can be captured, and executed successfully. Any test failures, exceptions and timeout will count as execution failure. 
\textit{Success Rate} Besides the execution rate, we also check whether the specification was satisfied. For Task II, we check whether the required target was in the generated code. For Task III, we check for code coverage. 

\textbf{Result}\quad
We report the Success Rate using $prompt_{full}$ in Table~\ref{tb:task23} and  $prompt_{part}$ in Table~\ref{tb:t2partial}. In the context free setting of this task, it provides \textit{no} context to the model, not even the ``file\_name'' required to complete the task. 
For most of the models, there is a significant performance boost from context-free to 8k. With only 8k context length, \claude{} achieves surprisingly high performance (46.2\%), 9.0\% ahead of the second best model \gpto{}. Some models, however, remain the same or even get slightly worse performance, including \codegemma{}, \codellama{}, \llamasm{}, and \gemini{}. 
The best performance is achieved by \claude{} and \gpto{} at 32k and 64k respectively. Starting at 16k, we have seen a sudden performance drop on \codellama{}, \mistral{}, \qwen{}, and \codestral{}. 

\section{Task III: Coverage-Oriented Test Implementation}
In this task, given some code blocks from source code, the model needs to complete the test code and cover those target blocks. This task shares a lot of similarity with Task II, so we will focus on the different part.

\textbf{Problem Formulation}\quad
For each problem, we provide $[s,f]$ and a specification to cover up to 10 code blocks from the source code. 
We provide the full code snippet in the ``Retrieved Code Snippets for Reference'' section of the prompt along with other oracle retrieval files. In the specification prompt, we provide the file name, the code blocks to cover, and the starting and ending line number for these code blocks in the original file. 

\textbf{Data Construction}\quad
We took a deterministic approach to select code blocks rather than randomly choosing code spans. 
We use the $Q_t^{peer}$ to guide the selection of code blocks to cover. 
For a case $y_t$, we check if there is some code blocks only covered by it, not any other peer cases. Typically it's the case where a conditional branch or a function is hit by only one case. 
We also filter out code blocks with fewer than 5 lines as we do not want to include many code fragments. 
The max number of files to cover is set at 10. 
With this approach, we can guide the model with a feasible and reasonable coverage requirement which also aligns with the function name and arguments.

The answer format and metrics of Task III remains the same with Task II. The different task specific prompt is shown in Figure~\ref{fig:prompt_box}. In this setting, since we include up to 10 files, we set the total sequence length to 64k and 128k to keep as many examples as possible.

\textbf{Results}\quad
We present the success rate of Task III in the rightmost 2 columns in Table~\ref{tb:task23}. \gpto{} achieves the best performance in 64k setting with 32.7\%. None of the open-source models passed 10\% in this task. \claude{} and \gptm{} are the models performing better with longer context provided. 

\textbf{Can LLMs satisfy the coverage requirement?} \quad
To answer this question, we compare the execution rate and the success rate in Table~\ref{tb:task3analyasis}. The gap of whether the coverage requirements can be met is around 5 to 10\% for different models. During evaluation, we only consider it a success if \textit{all} of the code blocks' coverage requirements are satisfied. As the result shows, there is generally a gap between execution rate v.s. success rate, indicates the generated test cases are generally not satisfying all the coverage requirements.

\input{tab_task3}

\input{related}

\section{Limitation \& Conclusion}

Our study is confined to Python and specific pytest tools. We didn't explore using code agents to tackle problems in this benchmark. Preliminary findings indicate that code agents, such as those by \citet{openhands}, are costly to run due to the need to explore entire repositories, primarily because of the overhead from reading large files.

In this study, we introduce a benchmark designed for multiple real-world software testing scenarios. We identified significant gaps in long-context handling, context utilization, and instruction-following capabilities between open-source and closed-source models, pointing to substantial opportunities for improvement. The coverage-driven oracle context could advance research on long-context evaluation in (code) LLMs. Researchers can use the API for real-time feedback to enhance models' coding and reasoning skills. Additionally, the data pipeline can be adapted to create high-quality training datasets.

\newpage
\input{impact}

\bibliography{my_bib}
\bibliographystyle{icml2025}


\appendix
\onecolumn

\section{Repositories Scrapped \& Used}
\label{app:sec:repo}

The benchmark incorporates the following repositories from GitHub: Pillow, elasticsearch-py, flask, httpx, jinja, kombu, paramiko, pip, requests, sqlalchemy, starlette, and pylint. Conversely, the repositories not utilized include: salt, celery, aiohttp, pytest, sphinx, docker-py, channels, mongoengine, boto3, scrapy, requests-html, black, dd-trace-py, ansible, pyzmq, python-prompt-toolkit, blessed, fastText, google-api-python-client, h2, scikit-learn, httpbin, ipython, libcloud, matplotlib, numpy, pandas, twisted, and voluptuous.

The cutoff date for this benchmark is set for August 30, 2024. Various reasons account for not using all repositories, such as:
\begin{itemize}
    \item some repositories do not support \texttt{pytest} and/or \texttt{pytest-cov},
    \item challenges in automatically configuring the environment or extracting test cases,
    \item non-standard naming of tests that disrupts our heuristics,
    \item failure of rule-based folder localization approach (a.k.a. finding \texttt{tests} and \texttt{src} folder) due to non-standard naming or project structure. For instance, \texttt{elasticsearch-py} has source code folder and test code folder named as \texttt{elasticsearch} and \texttt{test\_elasticsearch}. There are 11 repositories where we manually specified its folder name, 
    \item some repositories being very slow or causing issues during evaluation, and
    \item requirements for external setup, non-Python setup, or specific system configurations for some repositories.
\end{itemize}

\section{Results with $prompt_{part}$}
In Table~\ref{tb:t2partial} we present the results with $prompt_{part}$ on Task II. In Table~\ref{tb:task3part} we demonstrate the results on Task III. We found for most of the models, the $prompt_{full}$ which asks for the whole code block works better than $prompt_{part}$ in practice.

\input{prompt1}

\input{tab_task2_partial_block}

\begin{table}[t] 
\centering
\small
\begin{tabular}{@{}rcccc@{}} \toprule                                                 & \multicolumn{2}{c}{Success Rate} & \multicolumn{2}{c}{Execution Rate} \\ \cmidrule(l){2-5}  Model                                           & 64k         & 128k       & 64k         & 128k       \\ \midrule\zeroone               & 10.3\%      & 10.3\%     & 22.4\%      & 17.8\%     \\  \llamasm  & 0.9\%       & 0.9\%      & 4.7\%       & 1.9\%      \\ \llamamd & 11.2\%      & 4.7\%      & 14.0\%      & 7.5\%      \\\midrule \gptm                 & 21.5\%      & 20.6\%     & 29.9\%      & 28.0\%     \\ \gemini                   & 26.2\%      & 25.2\%     & 36.4\%      & 34.6\%     \\ \claude             & 32.7\%      & 29.9\%     & 42.1\%      & 37.4\%     \\ \gpto                 & 28.0\%      & 29.9\%     & 35.5\%      & 41.1\%     \\ \bottomrule \end{tabular} 
\caption{
Success rate and execution rate of Task III with $prompt_{part}$.
}
\label{tb:task3part}
\end{table}

\input{prompt2}
\input{prompt3}

\section{Prompt templates and examples}
\label{app:prompt}
We list the prompts for Task I in Fig~\ref{fig:task1po} and Fig~\ref{fig:task1ctx}, Task II in Fig~\ref{fig:task2prompt} and Task III in Fig~\ref{fig:task3prompt}.

\end{document}

%% file: tab_related_work_comp.tex
   \begin{table*}[!ht]
   \centering
   \scriptsize
   \setlength{\tabcolsep}{3pt}

   \begin{tabularx}{\textwidth}{r|p{3.5cm}|p{1.6cm}|p{1.0cm}|p{1.6cm}|p{0.7cm}|p{2cm}|p{1.4cm}} 
   \toprule
               & Use Case & Data Source & PL & Size & Exec Repo & Constructed Context & Coverage \\ \midrule
   SWE-bench \cite{swebench} & issue resolution & SWE-Bench & Python & 2294 & $\surd$ & Up to 50k &$\times$\\ 
   TestEval  \cite{testeval}  & test case gen & Leetcode & Python & 210 &$\times$& $\times$ & $\surd$ \\ 
   TestBench \cite{testbench}  & test case gen & Github & Java & 108 & $\surd$ &$\times$& $\surd$ \\ 
   SWT-Bench \cite{swtbench} & tcg for issue reproduction & SWE-Bench & Python & 1900+ & $\surd$ &$\times$& $\surd$ \\ 
   TestGenEval \cite{jain2024testgenevalrealworldunit} & test case gen & SWE-Bench & Python & 1210 & $\surd$ &$\times$& $\surd$ \\ \midrule
   CLOVER & test case gen (3 tasks covering completion \& generation) & Github (new) & Python & 845 from 16,234 & $\surd$ & Up to 128k & $\surd$ \\   \bottomrule
   \end{tabularx}
   \caption{Comparison of CLOVER and other benchmarks pertaining to test case generation. CLOVER encompasses three unique tasks for generating test cases. It includes 845 problems, leading to a total of 5312 instances when accounting for different context settings. Numerous benchmarks for test case generation are based on the work by \citet{swebench}.}
   \label{tb:related}
   \end{table*}

%% file: tab_task_stat.tex
\begin{table}[t]
\centering
\small
\setlength{\tabcolsep}{3pt}
\begin{tabular}{@{}ccccccccccc@{}}
\toprule
    & \#ex & avail & \# templ & PO & 4k  & 8k  & 16k  & 32k  & 64k  & 128k \\ \midrule
I   & 513 & 14952 & 1        & 0.9      & 2.7 & 5.5 & 11.0 & 21.9 & 42.8 & - \\
II & 151 & 184  & 2        & 1.0      & - & 3.9 & 11.8 & 27.7 & 58.7 & -  \\
III & 181 & 1098 & 2        & - & - & - & -  & -  & 57.7 & 93.5 \\ \bottomrule
\end{tabular}
\caption{Benchmark statistics. Number of unique ex(amples), number of templates and average number of tokens for different settings. Problem Only setting includes only the task instruction without supplementary context. One unique example will be populated into $n$ examples for various settings. For instance, in Task II, one unique example will be expanded into $2 \times 5=10$ examples.}
\label{tb:stat}
\end{table}

%% file: fig2.tex
\begin{figure}[t]
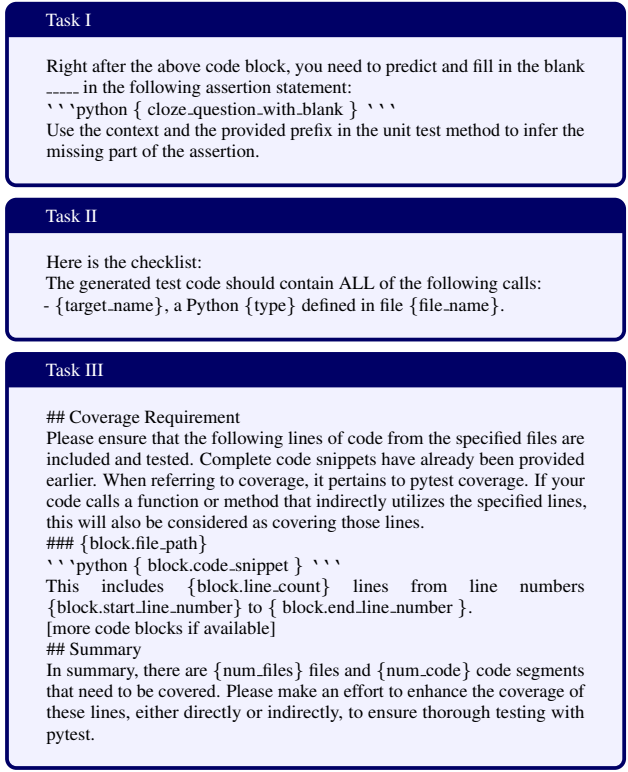

    \centering
    \scriptsize
    \begin{tcolorbox}[colback=blue!5!white, colframe=blue!40!black, title=Task I, width=\linewidth]
        
Right after the above code block, you need to predict and fill in the blank \_\_\_\_\_ in the following assertion statement:

\texttt{```}python
\{ cloze\_question\_with\_blank \}
\texttt{```}

Use the context and the provided prefix in the unit test method to infer the missing part of the assertion.

    \end{tcolorbox}
    \begin{tcolorbox}[colback=blue!5!white, colframe=blue!40!black, title=Task II, width=\linewidth]

Here is the checklist:
    
    The generated test code should contain ALL of the following calls:
    
    - \{target\_name\}, a Python \{type\} defined in file \{file\_name\}.

\end{tcolorbox}

    \begin{tcolorbox}[colback=blue!5!white, colframe=blue!40!black, title=Task III, width=\linewidth]
\#\# Coverage Requirement

Please ensure that the following lines of code from the specified files are included and tested. Complete code snippets have already been provided earlier. When referring to coverage, it pertains to pytest coverage. If your code calls a function or method that indirectly utilizes the specified lines, this will also be considered as covering those lines.

\#\#\# \{block.file\_path\}

\texttt{```}python
\{ block.code\_snippet \}
\texttt{```}

This includes \{block.line\_count\} lines from line numbers \{block.start\_line\_number\} to \{ block.end\_line\_number \}.

[more code blocks if available]

\#\# Summary

In summary, there are \{num\_files\} files and \{num\_code\} code segments that need to be covered. 
Please make an effort to enhance the coverage of these lines, either directly or indirectly, to ensure thorough testing with pytest.
    \end{tcolorbox}

    \caption{Task specific prompt template for Task I, II and III. For the complete prompts, check Sec~\ref{app:prompt}.}
    \label{fig:prompt_box}
\end{figure}

%% file: tab_task1_agg.tex
 \begin{table}[t] 
 
\centering
\small
  \begin{tabular}{lccc}                                   \toprule     & EM     & ER     & RER    \\ \midrule \codegemma                 & 30.3\% & 46.5\% & 43.7\% \\ \magicoder           & 23.3\% & 40.6\% & 34.3\% \\ \codellama   & 31.3\% & 52.6\% & 42.3\% \\ \starcoder   & 32.7\% & 52.4\% & 41.1\% \\ \mistral     & 21.5\% & 39.4\% & 37.2\% \\ \qwen        & 52.6\% & 64.7\% & 59.8\% \\ \codestral           & 45.5\% & 68.2\% & 63.3\% \\ \zeroone                 & 41.4\% & 61.7\% & 56.5\% \\ \llamasm  & 35.7\% & 57.4\% & 53.5\% \\ \llamamd & 49.9\% & 74.2\% & 69.0\% \\ 
  \midrule
  \gptm                 & 43.1\% & 70.8\% & 66.3\% \\ \gemini                   & 49.7\% & 71.9\% & 67.0\% \\ \claude             & 56.3\% & 78.2\% & 72.4\% \\ \gpto                      & 55.2\% & 72.0\% & 68.0\%  \\
  \bottomrule
  \end{tabular}
   \caption{Model performance on Task I in the Problem-Only setting. Refined execution rate (RER) is the recommended metric which reflects the models' ability in completing compilable and non-cheating assertion statements. Open source models are organized in ascending order based on their maximum supported length, followed by their model size. }
  \label{tb:task1}
  \end{table}

%% file: tab_task1_ctx.tex
\begin{table*}[ht] 
\centering
\footnotesize
\begin{tabular}{@{}cr|ccccccccc@{}} \toprule Max Len & Model &   PO &   4k &   8k &   16k &   32k &   64k &   Best &   $\Delta_{max} (\uparrow)$ &    $\Delta_{min} (\uparrow)$ \\ \midrule
8k & \codegemma &   43.7\% &   40.6\% &   41.5\% &  -  &   -  &  -  &   43.7\% &   -2.2\% &   -3.1\% \\
16k  & \magicoder &   34.3\% &   33.0\% &   34.2\% &   31.0\% &   -  &  -  &   34.3\% &   -0.1\% &   -3.3\% \\ 
16k  & \codellama &   42.3\% &   41.7\% &   42.3\% &   12.2\% &   -  &  -  &   42.3\% &   0.0\% &   \cellcolor[HTML]{F4CCCC}-30.1\% \\ 
16k  &  \starcoder &   41.1\% &   37.3\% &   25.0\% &   25.4\% &   -  &  -  &   41.1\% &   -3.8\% &   -16.1\% \\ 
32k & \mistral &   37.2\% &   34.1\% &   36.2\% &   12.2\% &   12.6\% &  -  &   37.2\% &   -1.0\% &   \cellcolor[HTML]{F4CCCC}-25.0\% \\ 
32k & \qwen &   59.8\% &   59.2\% &   60.4\% &   31.0\% &   30.7\% &  -  &   60.4\% &   \cellcolor[HTML]{D9EAD3}0.6\% &   \cellcolor[HTML]{F4CCCC}-29.1\% \\ 

32k &  \codestral &   63.3\% &   65.0\% &   64.2\% &   22.3\% &   22.1\% &  -  &   65.0\% &   \cellcolor[HTML]{D9EAD3}1.7\% &   \cellcolor[HTML]{F4CCCC}-41.2\% \\ \midrule
131k+ & \zeroone &   56.5\% &   55.6\% &   53.4\% &   43.9\% &   40.2\% &   40.6\% &   56.5\% &   -0.9\% &   -16.3\% \\ 131k & \llamasm &   53.5\% &   52.1\% &   51.3\% &   49.1\% &   49.0\% &   47.3\% &   53.5\% &   -1.4\% &   -6.2\% \\ 131k &  \llamamd &   69.0\% &   71.3\% &   71.1\% &   72.7\% &   71.7\% &   70.5\% &   72.7\% &   \cellcolor[HTML]{D9EAD3}3.7\% &   \cellcolor[HTML]{D9EAD3}1.5\% \\ 131k &  \gptm &   66.3\% &   66.0\% &   67.3\% &   65.3\% &   66.7\% &   67.7\% &   67.7\% &   \cellcolor[HTML]{D9EAD3}1.4\% &   -1.0\% \\ 131k+ & \gemini &   67.0\% &   67.6\% &   66.6\% &   66.0\% &   66.0\% &   66.2\% &   67.6\% &   \cellcolor[HTML]{D9EAD3}0.6\% &   -1.0\% \\ 131k & \claude &   72.4\% &   71.5\% &   73.4\% &   73.4\% &   74.4\% &   75.4\% &   75.4\% &   \cellcolor[HTML]{D9EAD3}3.0\% &   -0.9\% \\131k &  \gpto &   68.0\% &   71.0\% &   71.5\% &   70.9\% &   69.8\% &   67.2\% &   71.5\% &   \cellcolor[HTML]{D9EAD3}3.5\% &   -0.8\% \\  \bottomrule \end{tabular} 

\caption{Refined execution rate (RER) of models in contextual settings for Task I. The table lists models according to their maximum supported sequence length. The Problem-Only (PO) column indicates the baseline performance of models with average short input lengths (0.9k tokens). `Best' highlights the highest performance achieved across both settings (underscored for each model). Ctx Util measures the maximum and minimum performance change when shifting from baseline to contextual inputs.
We highlight $\Delta$ in green if $\Delta > 0\%$ and in red if $\Delta < -20\%$.
}
\label{tb:task1ctx}
\end{table*}

%% file: tab_task2.tex
\begin{table*}[t]
\footnotesize
\centering
\begin{tabular}{@{}c|cccccccc|cc@{}} \toprule 
\multirow{3}{*}{Model}                & \multicolumn{8}{c|}{Task II}                                              & \multicolumn{2}{c}{Task III} \\   &      &   \multicolumn{4}{c}{Performance at Max Seq Length} &   \multirow{2}{*}{Best} &   \multicolumn{2}{c|}{Ctx Util} &   \multicolumn{2}{c}{Perf} \\
 &      PO  & 8k     & 16k    & 32k    & 64k    &        &$\Delta_{max} $ \( \uparrow \)    & $\Delta_{min}$ \(\uparrow\)    & 64k         & 128k        \\ \midrule
 \codegemma                & \underline{14.9\%}  & 10.5\% &  -      &       - &      -  & 14.9\% & -4.4\% & -4.4\%  &       -      &       -      \\ 
 \magicoder &13.7\%	& \underline{20.2\%}	& 0.0\% &- &- & 20.2\%	& 6.5\% & 	-13.7\%  & -&- \\
 \codellama  & \underline{8.4\% }  & \underline{8.4\%}   & 0.0\%  &    -    &    -    & 8.4\%  & 0.0\%  & -8.4\%  &       -      &           -  \\ 
 \starcoder  & 13.8\% & \underline{19.1\%}  & 8.5\%  &    -    &     -   & 19.1\% & 5.3\%  & -5.3\%  &      -       &       -      \\ 
 \mistral    & 8.4\%  & \underline{11.6\%}   & 0.0\%  & 0.0\%  &     -   &  11.6\% & 3.2\%  & -8.4\%  &      -       &       -      \\ 
 \qwen       & 23.2\% & \underline{25.3\%}  & 0.0\%   & 0.0\%  &    -    & 25.3\% & 2.1\%  & \cellcolor{neg!30} -23.2\% &     -        &       -      \\ 
 \codestral          & 20.0\% & \underline{28.4\%}  & 0.0\%  & 0.0\%  &   -     & 28.4\% & 8.4\%  & \cellcolor{neg!30} -20.0\% &    -         &   -          \\ \midrule 
\zeroone                & 14.7\% & \underline{20.0\%}  & 14.7\% & 13.7\% & 13.7\% & 20.0\% & 5.3\%  & -1.0\%  & 4.7\%       & 3.7\%       \\ 
\llamasm & 5.3\%  & 5.3\%  & 4.2\%  &\underline{9.5\%}   & 5.3\%  & 9.5\%  & 4.2\%  & -1.1\%  & 1.9\%       & 0.9\%       \\ 
\llamamd &   12.9\% &   14.9\% &   13.8\% &  \underline{24.7\%}  &   18.3\% &   24.7\% &  \cellcolor{pos!30} 11.8\% &   0.9\% &   6.5\% &   3.7\% \\ 
\gptm                & 17.0\% & 22.3\% & 25.5\% & \underline{26.6\%}  & 25.5\% & 26.6\% & 9.6\%  & 5.3\%   & 19.6\%      &   21.5\%      \\ 
\gemini                  & 17.9\% & 17.9\% & \underline{23.2\%}  & 21.1\% & 16.8\% & 23.2\% & 5.3\%  & -1.1\%  & 28.0\%      & 27.1\%      \\ 
\claude            & 26.9\% & 46.2\% & 39.8\% & \underline{48.4\%}  & 46.2\% & 48.4\% &  \cellcolor{pos!30} 21.5\% &  \cellcolor{pos!30} 12.9\%  & 29.0\%      & 30.8\%      \\ 
\gpto                     & 28.7\% & 37.2\% & 42.6\% & 40.9\% & \underline{48.4\% } & 48.4\% &  \cellcolor{pos!30} 19.7\% & 8.5\%   & 32.7\%      & 31.8\%      \\ \bottomrule \end{tabular} 
\caption{Success rate on Task II and Task III (rightmost two columns). In Task II, we constructed the Problem-Only setting and contextual setting from 8k to 64k. In Task III, we only have the contextual setting since it heavily depends on retrieval files. For models that had a 0.0\% success rate,  with a manual inspection we discovered the outputs contained gibberish, such as backticks, line break symbols, and random words. We did not alter any configurations in vLLM for these models while testing with various lengths. 
}
\label{tb:task23}
\end{table*}

%% file: tab_task3.tex


\begin{table}[t] 
\centering
\footnotesize
\begin{tabular}{@{}rcccc@{}} 
\toprule                                                 & \multicolumn{2}{c}{Success Rate} & \multicolumn{2}{c}{Execution Rate} \\ \cmidrule(l){2-5}  Model                                           & 64k         & 128k       & 64k         & 128k       \\ \midrule
\zeroone            & 4.7\%       & 3.7\%      & 11.2\%      & 8.4\%      \\  \llamasm  & 1.9\%       & 0.9\%      & 2.8\%       & 4.7\%      \\ \llamamd & 6.5\%       & 3.7\%      & 8.4\%       & 10.3\%     \\ \midrule 
\gptm                 & 19.6\%      & 21.5\%     & 25.2\%      & 25.2\%     \\ \gemini                   & 28.0\%      & 27.1\%     & 38.3\%      & 36.4\%     \\ \claude             & 29.0\%      & 30.8\%     & 34.6\%      & 35.5\%     \\ \gpto                 & 32.7\%      & 31.8\%     & 42.1\%      & 39.3\%     \\ \bottomrule \end{tabular} 
\caption{Success rate and execution rate of models on Task III. }
\label{tb:task3analyasis}
\end{table}

%% file: related.tex
\section{Related Works}

\textbf{Code \& Test Case Generation Benchmarks}\quad
Developing evaluation benchmarks for code generation models is one of the most active research topics.
Earlier benchmarks like HumanEval \cite{chen2021evaluating} and MBPP \cite{austin2021program} focused on basic or LeetCode-style programming problems. BigCodeBench \cite{zhuo2024bigcodebench} extends this with complex function calls, while DevBench \cite{devbench} evaluates model performance in entire software development lifecycle. 
We summarized several recent benchmarks related to test case generation in Table~\ref{tb:related}. The source data and dev environment of \citet{swebench} has been widely adopted to develop new benchmarks. \citet{r2e} demonstrated a scalable framework to turn any GitHub repo into an interactive environment for agents.

\textbf{Test Case Generation}\quad
LLMs are widely used for automating test case generation. \citet{chatunitest} employs LLMs for efficient test creation. \citet{liu2024llm} utilize LLMs for bug detection, while \citet{tang2024chatgpt,yuan2024evaluating} enhance ChatGPT's unit test generation capabilities. \citet{alshahwan2024automated} explore LLM use in industry. Neural models for test case generation were proposed by \citet{tufano2020unit,nie2023learning}. \citet{ryan2024code} investigated coverage-guided test case generation with LLMs.

\textbf{Code LLMs and Agents}\quad
Recent studies on code-specific LLMs \cite{codellama,starcoder2,hui2024qwen2} showcase the potential of specialized models for code generation. Additionally, RepoAgent \cite{luo2024repoagent} proposes an agentic framework for repo-level documentation, SUPER \cite{bogin2024superevaluatingagentssetting} evaluates LLM-based agents on writing scripts and executing research tasks.

\textbf{Long Context for Code}\quad
Existing long context benchmarks either exclude coding tasks or have restricted access to code LLMs. RULER \cite{hsieh2024ruler} and $\infty$Bench \cite{zhang-etal-2024-bench} fail to replicate real-world software development scenarios. Meanwhile, code benchmarks are insufficient for long context evaluation. RepoBench \cite{repobench} sets the maximum token threshold of 12k for Python and 24k for Java. Most test cases in TestBench \cite{testbench} are within 32k tokens and TestGenEval \cite{jain2024testgenevalrealworldunit} evaluates context length up to 32k. SWE-Bench \cite{swebench} focuses on context length less than 50k, which is far behind many recent models often claiming to be able to consume 128k+ context length.




%% file: impact.tex
\section*{Impact Statements}

This paper presents research aimed at advancing the field of Machine Learning. There are a few potential societal impacts stemming from our work:

\paragraph{Safety Considerations}
While it is possible, though unlikely unless intentionally prompted, for LLMs to generate malicious or corrupted code, we disclaim responsibility for any consequences resulting from executing such code on our benchmark. 

We are committed to providing a safe and controlled evaluation environment by encapsulating our framework within a Docker container. We have implemented extensive precautionary measures to achieve this goal. However, in rare cases involving specific repositories that have been excluded from our benchmark, we have observed instances where the Docker container may exceed system memory limitations, potentially causing the host server to restart.

We advise users and researchers to carefully consider system configurations and setup when deploying any LLM-generated code directly on their machines.

\paragraph{Legal Compliance}
The usage of all repositories referenced in this paper is approved by the authors' organization following a thorough license check. The licenses include BSD-3-Clause, Apache-2.0, MIT, and LGPL-2.1.

%% file: prompt1.tex
\begin{figure}[t]
    \centering

\begin{tcolorbox}[colback=blue!5!white, colframe=blue!75!black,title=Task I (Problem Only)]

\textbf{Task Overview}

You need to complete the assertion in the provided unit test method by considering the context from the corresponding Python file.

\textbf{Context Information}

Below is the file context, which includes necessary imports and setups for the unittest:
\begin{verbatim}
{{ context }}
\end{verbatim}

\textbf{Target Unit Test Method}

This is the unit test method for which you need to complete the assertion statement:
\begin{verbatim}
{{ prefix }}
\end{verbatim}

\textbf{Your Task}

Right after the above code block, you need to predict and fill in the blank (\texttt{{cloze\_key}}) in the following assertion statement:
\begin{verbatim}
{{ cloze_question }}
\end{verbatim}
Use the context and the provided prefix in the unit test method to infer the missing part of the assertion.

\textbf{Output Format}

Your output should be a finished assertion statement. Do \textbf{not} generate the entire unit test method; only produce the assertion statement.

\textbf{Example}

For instance, if the cloze question is \texttt{assert \_\_\_\_\_\_ == (25, 25, 75, 75)}, and your prediction for the blank is \texttt{im.getbbox(alpha\_only=False)}, your output should be:

\begin{verbatim}
    assert im.getbbox(alpha_only=False) == (25, 25, 75, 75)
\end{verbatim}

Please provide your output in the desired format without additional explanations or step-by-step guidance.

\textbf{Notes}

\begin{itemize}
    \item The completed assertion should not be trivial. For instance, \texttt{assert True == True} and \texttt{assert str("a") == "a"} are considered trivial assertions.
    \item There will be precisely one blank in the assertion statement to be filled in.
\end{itemize}

\end{tcolorbox}

    \caption{PO prompt of Task I.}
    \label{fig:task1po}
\end{figure}

\begin{figure}[t]
    \centering

\begin{tcolorbox}[colback=blue!5!white, colframe=blue!75!black,title=Task I (Contextual)]

\textbf{Task Overview}

You need to complete the assertion in the provided unit test method by considering the context from the corresponding Python file.

\textbf{Retrieved Code Snippets for Reference}

Here are a few code snippets retrieved for your reference while making your prediction:

\begin{verbatim}
{{ retrieved_snippet }}
\end{verbatim}

\textbf{Context Information}

Below is the file context, which includes necessary imports and setups for the unittest:
\begin{verbatim}
{{ context }}
\end{verbatim}

\textbf{Target Unit Test Method}

This is the unit test method for which you need to complete the assertion statement:
\begin{verbatim}
{{ prefix }}
\end{verbatim}

\textbf{Your Task}

Right after the above code block, you need to predict and fill in the blank (\texttt{{ cloze\_key}}) in the following assertion statement:
\begin{verbatim}
{{ cloze_question }}
\end{verbatim}
Use the context and the provided prefix in the unit test method to infer the missing part of the assertion.

\textbf{Output Format}

Your output should be a finished assertion statement. Do \textbf{not} generate the entire unit test method; only produce the assertion statement.

\textbf{Example}

For instance, if the cloze question is \texttt{assert \_\_\_\_\_\_ == (25, 25, 75, 75)}, and your prediction for the blank is \texttt{im.getbbox(alpha\_only=False)}, your output should be:

\begin{verbatim}
    assert im.getbbox(alpha_only=False) == (25, 25, 75, 75)
\end{verbatim}

Please provide your output in the desired format without additional explanations or step-by-step guidance.

\textbf{Notes}
\begin{enumerate}
    \item The completed assertion should not be trivial. For instance, \texttt{assert True == True} and \texttt{assert str("a") == "a"} are considered trivial assertions.
    \item There will be precisely one blank in the assertion statement to be filled in.
\end{enumerate}

\end{tcolorbox}

    \caption{Contextual prompt of Task I.}
\label{fig:task1ctx}
\end{figure}

%% file: tab_task2_partial_block.tex
\begin{table}[] 
\centering
\small
\begin{tabular}{rcccccccc} 
\toprule
\multirow{2}{*}{}                      &        & \multicolumn{4}{c}{Performance at Max Seq Length} &        & \multicolumn{2}{c}{Ctx Util} \\                                      & PO & 8k     & 16k    & 32k    & 64k    & Best     &$\Delta_{max} $ \( \uparrow \)    & $\Delta_{min}$ \(\uparrow\)  \\ 
\midrule
\codegemma               & 10.6\%   & 11.7\% &   -     &   -     &   -     & 11.7\% & 1.1\%   & 1.1\%   \\ \magicoder         & 13.8\%   & 19.1\% & 0.0\%  & 0.0\%  & 3.2\%  & 19.1\% & 5.3\%   & -13.8\% \\ \codellama & 9.6\%    & 8.4\%  & 0.0\%  &   -     &   -     & 9.6\%  & -1.2\%  & -9.6\%  \\ \starcoder & 12.6\%   & 22.1\% & 9.5\%  &   -     &   -     & 22.1\% & 9.5\%   & -3.1\%  \\ \mistral   & 9.6\%    & 9.6\%  & 0.0\%  & 0.0\%  &   -     & 9.6\%  & 0.0\%   & -9.6\% \\ \qwen      & 27.4\%   & 20.0\% & 0.0\%  & 0.0\%  &   -     & 27.4\% & -7.4\%  & -27.4\% \\ \codestral         & 20.2\%   & 27.4\% & 0.0\%  & 0.0\%  &   -     & 27.4\% & 7.2\%   & -20.2\% \\ \midrule
\zeroone               & 12.6\%   & 23.2\% & 15.8\% & 6.3\%  & 7.4\%  & 23.2\% & 10.6\%  & -6.3\%  \\ \llamasm  & 7.4\%  & 11.6\%     & 6.4\%      & 9.5\%      & 9.6\%      & 11.6\% & 4.2\%         & -1.0\%       \\ \llamamd & 14.0\% & 12.9\%     & 14.9\%     & 13.7\%     & 12.8\%     & 14.9\% & 0.9\%         & -1.2\%       \\ \gptm               & 14.7\%   & 27.7\% & 22.1\% & 24.5\% & 20.2\% & 27.7\% & 13.0\%  & 5.5\%   \\ \gemini                 & 20.0\%   & 21.1\% & 26.3\% & 24.2\% & 23.2\% & 26.3\% & 6.3\%   & 1.1\%   \\ \claude             & 28.4\% & 45.3\%     & 44.2\%     & 46.3\%     & 48.4\%     & 48.4\% & 20.0\%        & 15.8\%       \\ \gpto                    & 25.8\%   & 33.3\% & 39.8\% & 39.4\% & 44.7\% & 44.7\% & 18.9\%  & 7.5\%  \\  \bottomrule

 \end{tabular} 
 \caption{Success Rate of Task II with $prompt_{part}$. }

 \label{tb:t2partial}
 \end{table}

%% file: prompt2.tex
\begin{figure}
    \centering
    \scriptsize
    \begin{tcolorbox}[colback=blue!5!white, colframe=blue!75!black]

\textbf{Task Overview}

Create a unit test based on the provided method signature and given context. All necessary imports for the test have been included, so focus solely on writing the unit test method. No additional libraries can be imported.

\textbf{Retrieved Code Snippets for Reference}

Several code snippets have been provided for your reference:
\begin{verbatim}
{{ retrieved_snippet }}
\end{verbatim}

\textbf{Hint:} Depending on the target unit test method's signature, you might want to utilize or test these code snippets. 

\textbf{Requirement:} Each retrieved code snippet must be used at least once in the final unit test method.

\textbf{Context Information}

Below is the test code context, including necessary imports and setups for unittest:
\begin{verbatim}
{{ context }}
\end{verbatim}

\textbf{Signature of Target Unit Test Method}

This is the unit test method signature you need to complete:
\begin{verbatim}
{{ prefix }}
\end{verbatim}

\textbf{Your Task}

Continue writing the unit test method immediately following the provided method signature, ensuring to include at least one substantial assertion. Your task is to replace ``$<$YOUR CODE HERE$>$'' with correctly indented code.

\textbf{Output Format}

Please return the \textbf{WHOLE} \textbf{FILE} content (including the context and the completed unit test method, but not the Retrieved Code Snippets). You should not modify the context part of the code.

\textbf{Example 1:}
\begin{verbatim}
class A:
    def __init__(self):
        self.a = 1
        self.b = 2
    def test_comparing_a_b():
<YOUR CODE HERE>
    def test_assert_b():
        assert self.b == 2
\end{verbatim}

The output should follow the correct indentation principles. For example:
\begin{verbatim}
class A:
    def __init__(self):
        self.a = 1
        self.b = 2
    def test_comparing_a_b():
        assert self.a == 1       # Correct indentation with eight spaces before `assert`
        assert self.a != self.b  # Correct indentation with eight spaces before `assert`
    def test_assert_b():
        assert self.b == 2
\end{verbatim}

Here is one WRONG example:
\begin{verbatim}
# wrong example
class A:
    def __init__(self):
        self.a = 1
        self.b = 2
    def test_comparing_a_b():
    assert self.a == 1       # Incorrect indentation
assert self.a != self.b  # Incorrect indentation
    def test_assert_b():
        return           # YOU SHOULD NEVER modify the context code
\end{verbatim}

\textbf{Example 2:}
\begin{verbatim}
def test_value_c():
<YOUR CODE HERE>
\end{verbatim}

The output should follow the correct indentation principles. For example:
\begin{verbatim}
def test_value_c():
    assert c == 3
\end{verbatim}

- Keep in mind that the ``$<$YOUR CODE HERE$>$'' section lacks any leading spaces. 
- Ensure you include the appropriate amount of indentation before the `assert` statements. 
- If the method signature necessitates returning an object, implement that as well.

\end{tcolorbox}

    \caption{Contextual $prompt_{full}$ of Task II.}
    \label{fig:task2prompt}
\end{figure}

%% file: prompt3.tex
\begin{figure}
    \centering
    \small

\begin{tcolorbox}[colback=blue!5!white, colframe=blue!75!black]

\textbf{Task Overview}

Create a unit test based on the provided method signature and given context. All necessary imports for the test have been included, so focus solely on writing the unit test method. No additional libraries can be imported.

\textbf{Retrieved Code Snippets for Reference}

Several code snippets have been provided for your reference:

\begin{verbatim}
{{ retrieved_snippet }}
\end{verbatim}

\textbf{Output Format}

When completing a code snippet, the output should include the entire code context provided, including any imports, class definitions, or function signatures. Replace placeholders with meaningful code that fits the context. Example:

\textbf{Given:}
\begin{verbatim}
import pytest
from aiohttp import web
from aiohttp.web_urldispatcher import UrlDispatcher

@pytest.fixture
def router() -> UrlDispatcher:
    return UrlDispatcher()

def test_get(router: UrlDispatcher) -> None:
{{ SYMBOL_YOUR_CODE }}
\end{verbatim}

The goal is to complete the method \texttt{test\_get(router: UrlDispatcher) -> None:} with correct indentation and necessary logic while retaining the full context.

\textbf{Example output:}
\begin{verbatim}
import pytest
from aiohttp import web
from aiohttp.web_urldispatcher import UrlDispatcher

@pytest.fixture
def router() -> UrlDispatcher:
    return UrlDispatcher()

def test_get(router: UrlDispatcher) -> None:
    async def handler(request: web.Request) -> NoReturn:
        assert False

    router.add_routes([web.get("/", handler)])
    route = list(router.routes())[1]
    assert route.handler is handler
    assert route.method == "GET"
\end{verbatim}

\textbf{Task with Context Provided}

Below is the test code context, including necessary imports and setups for \texttt{unittest}. Following the method signature "\texttt{{ method\_signature }}", you need to complete the unit test method. Ensure to include at least one substantial assertion. Replace \texttt{{ SYMBOL\_YOUR\_CODE }} with the correctly indented test code.

\begin{verbatim}
{{ context }}
\end{verbatim}

\textbf{Hint:} Depending on the target unit test method's signature, you might want to utilize or test these code snippets. 

\textbf{Requirement:}
\begin{itemize}
  \item Ensure you return the unit test method including the given method signature. Do not modify the method signature.
  \item You are not allowed to import anything else as all necessary imports for the case have been provided.
  \item Properly indent each line before including it.
  \item Avoid using ANY trivial assertions such as \texttt{assert True == True} or \texttt{assert str("a") == "a"}, as they will be deemed incorrect.
\end{itemize}

\begin{verbatim}
{{ coverage_requirement }}
\end{verbatim}

\end{tcolorbox}
    \caption{Contextual $prompt_{full}$ for Task III. }
    \label{fig:task3prompt}
\end{figure}